\documentclass[aps, prl,twocolumn, superscriptaddress, preprintnumbers, floatfix, longbibliography, 10pt]{revtex4-1}

\usepackage{amssymb}
\usepackage{soul}
\usepackage{graphicx}
\usepackage{color}
\usepackage{comment}
\usepackage[colorlinks,allcolors=blue]{hyperref}
\usepackage{subfigure}
\usepackage{amsmath}
\usepackage{newtxtext}
\usepackage{newtxmath}
\usepackage{microtype}
\usepackage{bm}
\usepackage[per-mode=symbol]{siunitx}
\usepackage[capitalize]{cleveref}

\usepackage[colorlinks,allcolors=blue]{hyperref}
\usepackage[capitalize]{cleveref}

\definecolor{Red}{rgb}{0.8,0,0}
\definecolor{DarkGreen}{rgb}{0,0.8,0}
\definecolor{Pink}{rgb}{1,0,1}
\definecolor{Purple}{rgb}{0.7,0,1}
\definecolor{Orange}{rgb}{1,0.6,0}
\definecolor{Teal}{rgb}{0,0.5,0.5}
\definecolor{Blue}{rgb}{0,0,1}

\begin{document}

\title{Large tunable thermoelectric effects in superconducting spin valves \\with commercially available materials}

\author{Pablo Tuero$^{\dag}$}
\affiliation{Departamento F\'isica de la Materia Condensada C-III, Universidad Aut\'onoma de Madrid, Madrid 28049, Spain}

\author{Johanne Bratland  Tjernshaugen $^{\dag}$}
\affiliation{Center for Quantum Spintronics, Department of Physics, Norwegian University of Science and Technology, NO-7491 Trondheim, Norway}

\author{Carlos Sanchez}
\affiliation{Departamento F\'isica de la Materia Condensada C-III, Universidad Aut\'onoma de Madrid, Madrid 28049, Spain}

\author{César Gonzalez-Ruano}
\affiliation{Departamento F\'isica de la Materia Condensada C-III, Universidad Aut\'onoma de Madrid, Madrid 28049, Spain}
\affiliation{Department of Electrical Engineering and Institute for Research in Technology, ICAI School of Engineering, Comillas Pontifical University, C/Alberto Aguilera 25, 28015 Madrid, Spain}

\author{Yuan Lu}
\affiliation{Université de Lorraine, CNRS, IJL, F-54000 Nancy, France}

\author{Jacob Linder}
\email[e-mail: ]{jacob.linder@ntnu.no}
\affiliation{Center for Quantum Spintronics, Department of Physics, Norwegian University of Science and Technology, NO-7491 Trondheim, Norway}

\author{Farkhad G. Aliev}
\email[e-mail: ]{farkhad.aliev@uam.es}
\affiliation{Departamento F\'isica de la Materia Condensada C-III, Instituto Nicol\'as Cabrera (INC) and  Condensed Matter Physics Institute (IFIMAC), Universidad Aut\'onoma de Madrid, Madrid 28049, Spain}

\begin{abstract}
Recent studies have revealed magnetically controllable thermoelectric effects in superconductor/ferromagnet (S/F) structures. A tunable cryogenic thermoelectric generator needs not only a high conversion factor between electricity and heat, but also a large change in the thermoelectric output when switching the magnetic state of the device.
However, the reported modifications in thermoelectric power are either minimal, involve superconductors with relatively low critical temperatures (below 1 K), or do not utilize commercially available spintronic materials. Here, we experimentally measure and numerically model thermoelectric effects in fully epitaxial F/S/F junctions based on commercially available, easily grown materials, as well as their dependence on the magnetic configuration of the F electrodes. We observe sizeable Seebeck coefficients for the parallel alignment of the ferromagnetic electrodes, reaching values of about $100$~$\mu$V/K. Importantly, we find a decrease of the thermoelectric signal of more than an order of magnitude when switching from a parallel to an antiparallel configuration, constituting a large thermoelectric spin-valve effect. Theoretical modeling based on a self-consistent non-equilibrium Keldysh-Usadel Green's function theory, combined with micromagnetic simulations, qualitatively reproduce the experimental findings.  The thermoelectric effect is optimized when there is a large spin-dependent electron-hole asymmetry in the superconductor combined with spin-dependent transmission through the interfaces. 
These findings pave the way for the development of efficient and versatile cryogenic thermoelectric heat engines. 
\end{abstract}

\maketitle

\section{Introduction} 

The conversion of heat into electricity and vice versa is known as thermoelectricity \cite{benenti2017fundamental, bergeret_rmp_18}. This field of research has received increasing attention over the last decades, not only for the purpose of gaining insight into the fundamental physics of new materials~\cite{aliev_zpbc_90,Ojha2024}, but also due to the large number of industrial applications thermoelectricity enables. These include vastly different areas such as temperature control (cooling) of electronic devices, air conditioning, food refrigeration, and ultrasensitive detection of electromagnetic radiation~\cite{tritt_armr_11,wei_jms_20,soleimani_seta_20,shi_chemrev_20,irfan_mset_24}.

A prominent challenge with thermoelectricity occurs in the low-temperature regime. Whereas conventional thermoelectric materials typically operate in the regime 100~K -- 400~K, their performance, in terms of for instance the Seebeck coefficient, drastically drops at lower temperatures~\cite{feng_nanoenergy_24}. Therefore, there exists a need to design better material platforms that can yield efficient thermoelectricity even under cryogenic conditions. One promising application of thermoelectricity under cryogenic conditions is localized cryogenic cooling. This can be achieved using either thermoelectric materials highly efficient at low temperatures~\cite{harutyunyan_apl_03} or superconductor based microscale refrigerators \cite{Muhonen2012} which would complement the existing liquid nitrogen, helium or magnetic salt cooling procedures.

Superconductors are excellent thermoelectric materials when combined with ferromagnets~\cite{bergeret_rmp_18, heikkila_pss_19}. This emergent phenomenon has been predicted {more than a decade ago}~\cite{machon_prl_13,ozaeta_prl_14} and subsequently experimentally confirmed~\cite{kolenda_prl_16}. The key mechanism behind this emergent thermoelectricity is a spin-dependent particle-hole asymmetry in a superconductor that coexists with a spin-splitting field, which in turn is utilized by using the spin-selective transport properties of ferromagnets. Combining superconductors with antiferromagnets also produces enhanced thermoelectric effects~\cite{sukhachov_prb_24}. Moreover, large thermoelectric effects have been predicted \cite{marchegiani2020nonlinear} and experimentally confirmed \cite{germanese2022bipolar} in tunnel junctions with two different superconductors in the non-linear regime due to spontaneous breaking of electron-hole symmetry.

 The investigation of heat dissipation and energy harvesting in superconductor/ferromagnet (S/F) hybrids and in quantum devices in general \cite{meng2024} is an important area of research. For example, quantum error correction operations produce heating \cite{bilokur2024Arxiv}, 
potentially harvestable by low-temperature Peltier elements.
 In that context, obtaining the maximum possible Seebeck coefficients at the lowest possible temperatures (where actual quantum computers operate) and with minimum possible temperature gradients, could be important for the potential applications of magnetic state controlled thermoelectricity.
Magnetic control over thermoelectric effects could improve localized cooling for quantum devices, such as superconducting qubits and ultra-sensitive detectors \cite{Heikkil2018} by dynamically tuning heat flow. Magnetically controlled thermoelectric transport could also be used to manipulate heat flux (thermal diodes) \cite{Rajapaksha2024}, in a way analogous to how electronics manipulate electric energy. Last but not least, the above-mentioned waste heat from superconducting or other low-temperature electronic systems could be converted into usable electrical energy via magnetically tunable thermoelectric generators \cite{heikkila_pss_19, araujo_natcom_24}.

Very recently, it has been predicted~\cite{ouassou_prb_22} and experimentally demonstrated~\cite{gonzalez_prl_23,araujo_natcom_24} that interfacing a superconductor with two ferromagnets allows the Seebeck coefficient of the system to be tuned via the magnetic alignment of the ferromagnetic regions. However, in order for such an effect to be viable for cryogenic thermoelectric applications, a number of conditions have to be met. Firstly, one would need to use commercially available materials with high spin polarization which are easily grown. Secondly, the resulting magnitude of the thermoelectric effect would need to be sizeable at temperatures well below $T_c$, and there should be a large difference in the Seebeck coefficient for the parallel (P) and antiparallel (AP) magnetic states of the spin-valve.

In this work, we report the experimental observation of a large superconducting thermoelectric spin-valve effect that meets all of the key criteria stated above. This represents a major advance compared to previous works on related superconductor/ferromagnet structures \cite{gonzalez_prl_23,araujo_natcom_24}. Specifically, we study the spin-dependent transport and thermoelectric response in Fe/MgO/V/MgO/Fe/Co structures as a function of the relative alignment between the soft (Fe) and hard (Fe/Co) F electrodes. 
An important reason to use vanadium in our Fe/MgO/V based junctions, is its nearly perfect crystalline structure matching with Fe and MgO, allowing the superconductor to be grown on top of an almost fully spin polarized (Fe/MgO) ferromagnet with effective spin polarization exceeding $80\%$.
Our findings demonstrate a large thermoelectric effect, with a Seebeck coefficient exceeding $100~\mu$V/K at base temperatures around $0.1T_c$, where $T_c$ is the critical temperature of the V film. Importantly, we report a large change in the thermoelectric response -- greater than a factor of 10 -- when switching the device from the antiparallel (AP) to the parallel (P) alignment of the electrodes. Analyzing the results, we find a high sensitivity of the thermoelectric response to magnetic domain rotation and motion, as highlighted by the comparison of experimental thermoelectric and magnetoresistance data with micromagnetic simulations and theoretical modeling based on the non-equilibrium Keldysh-Usadel Green's function formalism.

\begin{figure}
\begin{center}
\includegraphics[width=\linewidth]{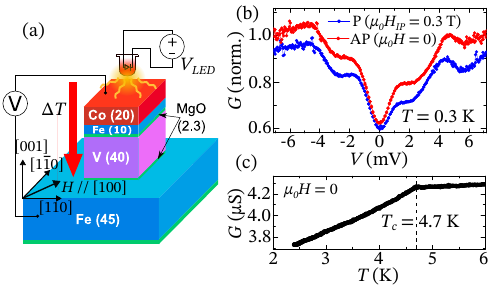}
\caption{(a)~Sketch of the F/S/F junctions when heated by a light emitting diode (LED). The black arrows show the crystalline axes and the direction of the applied magnetic field. The layer thicknesses are given in nm. (b)~Conductance--bias curves under different magnetic configurations measured at $T=0.07T_c$. Conductance is normalized so that $G(V=7 \text{ mV})=1$ when $\mu_{0}H=0$. (c)~Temperature dependence of the zero bias conductance measured with a current bias $I_{bias}=\pm0.5$~nA with $\mu_{0}H=0$, showing a superconducting critical temperature of 4.7~K indicated by the dashed line. The decrease in conductance below $T_c$ corresponds to the opening of a pseudogap in the conductance due to superconductivity.} 
\label{Fig1}
\end{center}
\end{figure}

\section{Experimental results}
\Cref{Fig1}(a) illustrates the experimental setup and the specific junctions under study. We measure magnetoresistance (MR) and thermoelectric (TE) effects in epitaxial Fe(45)/MgO(2.3)/V(40)/MgO(2.3)/Fe(10)/Co(20) junctions grown on MgO(001) substrates, with layer thicknesses provided in nanometers. Further details on junction growth, characterization, experimental setup, and simulation procedures are available in the Supplemental Material~\cite{supplemental}. 
In these F1/S/F2 junctions, V is the BCS superconductor (S), Fe serves as the soft magnetic electrode (F1) and Fe/Co functions as the magnetically hard electrode (F2), while the MgO acts as a symmetry-filtering tunnel barrier, facilitating high spin polarization ($p$) in Fe, exceeding 0.7-0.8~\cite{Martinez2018,GonzalezRuano2020}. Due to the weak antiferromagnetic coupling of the junctions~\cite{kochelaev_zetf_79,khusainov_zetf_96, DiBernardo2019, Tuero2024}, two distinct magnetic states are experimentally observed: at low fields, the junctions tend toward an antiparallel alignment, while a moderate external magnetic field ($H$) of 0.1–0.2~T will drive them into a parallel magnetic state.
\Cref{Fig1}(b) provides a low-bias electron transport characterization of the junctions in the superconducting state for both magnetic alignments. The oscillations in the conductance curves are likely due to McMillan resonances \cite{Tuero2024,rowell1966electron} that may occur for clean samples. \Cref{Fig1}(c) provides temperature-dependent subgap conductance measurements that show a superconducting critical temperature of $T_{c}=4.7$~K marked by an abrupt change of slope. This value is close to the one expected from the dependence of $T_c$ on the thickness for vanadium thin films grown in ultra high vacuum conditions \cite{Alekseevskii1976}. The steeper decrease in conductance at biases less than the superconducting gap and below $T_{c}$ is due to the reduced density of states of vanadium at the Fermi level in the superconducting state. An alternative method to determine the superconducting critical temperature, which provides similar results, is related to the detection of the emergence of superconducting quasiparticle interference effects in conductance as a function of temperature \cite{Tuero2024}.

In order to perform thermoelectric response measurements, we induce temperature gradients across the junctions by controlling the power dissipated by a light emitting diode (LED) placed above them. This creates an inward heat flux on the top surface of the samples that can be tuned by biasing the LED at different voltages $V_{LED}$. We estimate the temperature profile and the total temperature difference $\Delta{T}$ across the heterostructure by solving the heat diffusion equation in a one-dimensional model that approximates our samples \cite{supplemental}. We note that the estimated values of $\Delta{T}$ are in the order of $0.1-0.2$~K, which indicates that the device could operate in slightly nonlinear regime in measurements with a base temperature of 0.3 K.In the absence of applied current ($I=0$), the TE voltage $\Delta{V}$ generated under a given $\Delta{T}$ is obtained by subtracting the background voltage signal from the voltage measured under heating (see \cite{supplemental} for the discussion of the origin of the background TE signal).

\begin{figure}
\begin{center}
\includegraphics[width=\linewidth]{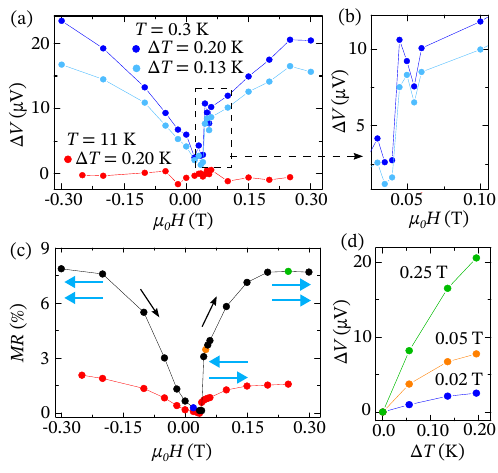}
\caption{Thermoelectric response of a F/S/F junction as a function of the upward swept applied in-plane (IP) magnetic field at $T=0.3$~K and two different evaluated $\Delta T$ values. For comparison, a much smaller thermoelectric response observed above the critical temperature (at $T=11$~K) is also shown. Part (b) zooms the low field region in (a), where an abrupt transition from the AP to the P state takes place. (c) Typical magnetoresistance (MR) curves vs IP magnetic field at a bias of $V=5$~mV at $T=0.3$~K and $11$~K. The blue arrows show the relative magnetic alignment of the ferromagnetic electrodes while the black arrow shows the direction of variation of the magnetic field. (d) Dependence of the TE response on the evaluated temperature difference between the ferromagnetic electrodes, measured at different IP magnetic states as indicated by the colored points in (c).}
\label{Fig2}
\end{center}
\end{figure}

\Cref{Fig2}(a) presents our main experimental observation: a substantial (over $10\times$) change in $\Delta{V}$ between the AP and P alignments of the ferromagnetic electrodes. The Seebeck coefficient for the parallel alignment, calculated based on the estimated temperature gradients, reaches a 
value of approximately $100~\mu$V/K~\cite{supplemental}. The strong magnetic field dependence of the TE signal disappears above $T_c$.
The data shown in Figure \ref{Fig2} corresponds to an upward sweep of the external in-plane magnetic field. Starting from the most negative $H$, as its magnitude
decreases and the magnetic configuration transitions from the P to AP state, the TE signal changes relatively smoothly. In contrast, the transition from AP to P in the positive $H$ branch leads to an abrupt increase in the TE response, accompanied by a subtle but reproducible ``overshoot'' effect manifested as a peak followed by an immediate drop, shown in detail in \Cref{Fig2}(b).

Magnetoresistance (MR) measurements performed during the TE experiments are shown in \Cref{Fig2}(c). Resistance $R$ is measured at a bias of $V\sim5$~mV under each value of applied in-plane magnetic field $H$ in thermal equilibrium before heating the junction with the LED. We define the magnetic field dependent MR as:
\begin{equation}\label{eq:MR}
    {MR}(H)=\frac{R(H)-R_{min}}{R_{min}}\times100\%.
\end{equation}
Here, $R_{min}$ is the minimum measured value of resistance for a given base temperature, and therefore corresponding to the magnetic state closest to a fully AP state.

Interestingly, we find that resistance is higher in the P state than in the AP state, contrary to the seminal Fe/MgO/Fe-based heterojunctions 
\cite{gonzalez_prl_23,Parkin2004:NM,Yuasa2004:NM}. We believe that the main reason for the negative MR in our F/S/F (F/N/F above $T_{c}$) junctions is that  two Fe layers are separated 
by a relatively thick 40 nm vanadium film.
Therefore, the simple picture explaining the conventional tunneling MR 
in Fe/MgO/Fe junctions may not be applicable to our structure. Importantly, early works on tunneling MR demonstrated that the presence of different non-magnetic spacers in between two ferromagnetic electrodes can alter not only its magnitude but also its sign \cite{DeTeresa1999}. More recently, the introduction of a vanadium film between Fe and MgO/Fe has been predicted to invert the MR sign \cite{Sanvito2009}.

We observe that the MR curve remains qualitatively similar above and below $T_c$, i.e., it decreases more smoothly in the P to AP transition in the negative $H$ branch and increases more abruptly 
in the AP to P transition in the positive $H$ branch. As we discuss below, based on micromagnetic simulations, the difference between P-AP and AP-P transitions
could be due to a rapid change in the relative magnetization configuration of the F electrodes during AP-P transition due to the combined effects of macrospin rotation and domain wall formation and displacement. 
The overall magnitude of the MR increases by approximately a factor of three in the superconducting state. We believe that the enhanced negative MR observed well below $T_c$ could be attributed to a more effective spin signal transfer between the two Fe electrodes through the 40 nm thick V layer due to the generation of superconducting spin-triplet states \cite{martinez_prb_20, GonzalezRuano2020}.

\Cref{Fig2}(d) illustrates the TE response as a function of the temperature gradient at various fixed magnetic fields, color-coded to correspond with the MR plot in \Cref{Fig2}(c), confirming a strong dependence of the TE response with the magnetic state of the junctions.

\begin{figure*}
\begin{center}
\includegraphics[width=0.94\linewidth]{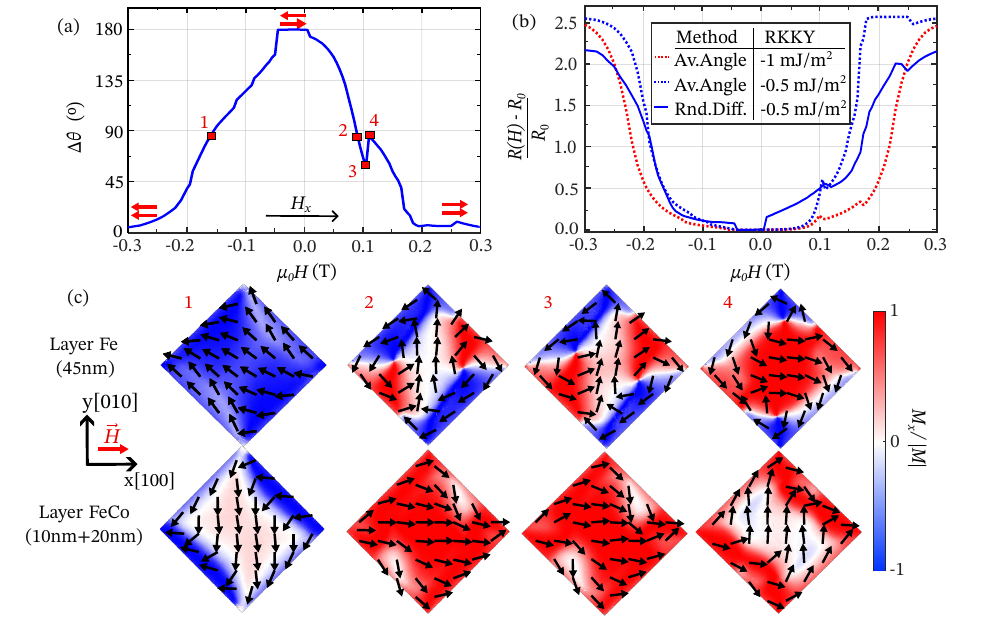}
\caption{(a) Angle difference between the macrospin of the two F regions for an AF coupling of RKKY $=-0.5$~mJ/m$^2$ (selected points pointed with numbers). (b) Normalized magnetoresistance for two different couplings and calculated using two methods explained in the text, where $R_0$ is the resistance at $\mu_{0}H=0$. (c) Magnetization map for the interface of both F layers on the selected points in (a).}
\label{Fig3}
\end{center}
\end{figure*}

\section{Micromagnetic simulations}\label{sec:micromagnetic simulations}

Numerical simulations of the TE response require knowledge about the average angle between the magnetizations of the two ferromagnets enclosing the superconductor. This information was obtained through micromagnetic simulations using the open-source software MuMax$^3$~\cite{Mumax2014}. 
At the same time, in order to mimic the experimental setup of \Cref{Fig1}(a), the system was divided into three different regions. Region 1 models the soft 45~nm thick Fe layer, while region 2 is the 10~nm thick Fe layer and Region 3 the 20~nm thick Co layer, that together form the hard Fe/Co electrode (see \cite{supplemental} for simulation details). To simulate a more realistic system, the presence of defects in the crystalline structure of the layers was taken into account. A concentration of 25\% for superficial and 10\% for bulk defects was considered.~\cite{supplemental}.

In our epitaxial Fe/MgO/V/MgO/Fe/Co vertically patterned junctions grown on MgO(001) substrate, the Fe(001) surface follows the cubic symmetry of the underlying MgO(001) lattice, leading to a dominant fourfold anisotropy with two easy axes along the two diagonals of the rectangular patterned junctions (along the MgO sides). This stabilizes two equivalent Fe (100) and Fe (010)  magnetization directions marked (x) and (y) in Figure~\ref{Fig3}(c) indicating both equivalent directions. As sketched in Figure~\ref{Fig1}(a), Fe/Co are two separate layers, i.e. not intermixed Fe-Co. The first Fe(10nm) layer interfacing MgO has a fourfold crystalline anisotropy, similarly to the soft Fe layer, and is hardened by the Co(20nm) layer. The micromagnetic simulations did not assign any in-plane or out-of-plane anisotropy to the Co layer to simulate magnetization reorientation and magnetoresistance. The possible out of plane anisotropy is irrelevant in our case as the Co layer thickness substantially exceeds 1-2 nm, where perpendicular magnetic anisotropy has been observed \cite{Dieny2017}.

Moreover, owing to the presence of the low field antiferromagnetic (AF) coupling between the two F layers, a negative exchange interaction was introduced between regions 1 and 2 by means of a negative Ruderman–Kittel–Kasuya–Yosida (RKKY) coupling constant. Different RKKY constants from $-0.5$~mJ/m$^2$ to $-2$~mJ/m$^2$ were considered.
The simulations were performed by varying the magnetic field starting from $-0.5$~T and up to $0.5$~T. The results for the angle difference $\Delta\theta(H)$ in \Cref{Fig3}(a) and \Cref{fig:numerical_model}(c) were computed based on the macrospin of each region. In other words, the average magnetization vector for AF coupled regions 1 and 2 (see Fig.S5 in ~\cite{supplemental})  was computed separately, and then the angle between them was calculated.

To calculate the MR we followed a simplified model~\cite{PhysRevB.39.6995} to approximate the magnetic state dependent conductance $G$ for our junctions (see \cite{supplemental} for details). We simulated the MR using two approaches. The first one (Av.Angle) calculates conductance as a function of the angle of the averaged magnetization in the two layers. The second approach divides the magnetic layers into a number of interfacial cells and then averages the conductance between each cell (discretization size of $2.34\times2.34\times1.67$~nm$^3$ was used, see \cite{supplemental} for the details) from the one electrode and a randomly chosen cell of the opposite electrode. Then, the electrodes were interchanged and same procedure was repeated. The second method (Rnd.Diff.) thus models elastic impurity scattering of the electrons within the 40nm thick V  
layer separating the ferromagnetic electrodes. As one observes in \Cref{Fig3}(b), using the lowest AF coupling value of RKKY$=-0.5$~mJ/m$^2$, the above mentioned concentration of defects and the Rnd.Diff. method, we obtain an MR curve that is qualitatively similar to the experimental measurements in \Cref{Fig2}(c). The magnetoresistance saturates above
0.3 T and the ``overshoot'' effect is reduced as in MR experiments.

\section{Thermoelectricity via non-equilibrium quasiclassical theory}

We use non-equilibrium Keldysh-Usadel Green's function theory \cite{bergeret_rmp_18, theta_parameterization} to numerically explore the setup in Figure \ref{Fig1}(a). We first provide an overview of the model used. A more technical description of the modeling is provided in \cite{supplemental}. Second, we combine the results from the quasiclassical simulations with the micromagnetic simulations to model how the thermoelectricity depends on the external magnetic field, and we compare the result with the experimental data in Figure \ref{Fig2}(a).

\begin{figure}
\begin{center}
\includegraphics[]{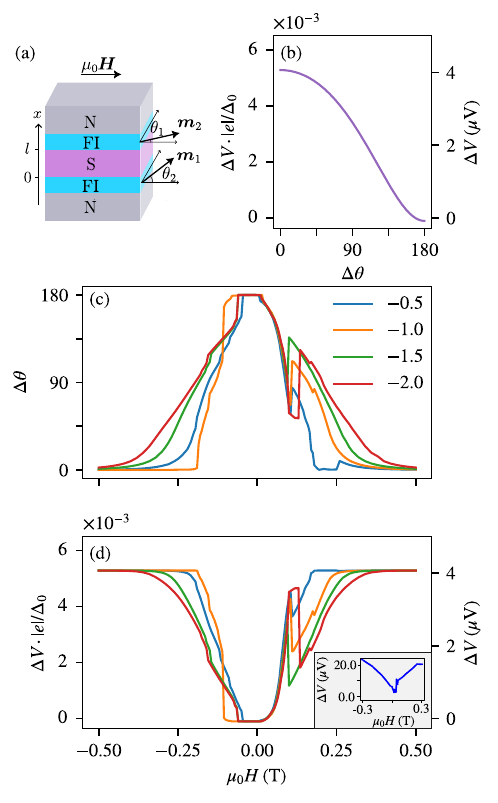}
\caption{(a) The model used in the numerical calculations. The superconductor S is sandwiched between two ferromagnets, modeled as non-superconducting reservoirs N with spin-active interfaces FI. The angle difference between the average in-plane magnetizations $\boldsymbol{m}_1$ and $\boldsymbol{m}_2$ of the interfaces is given by $\Delta\theta=\theta_2-\theta_1$. The external magnetic field $\mu_0\boldsymbol{H}$ affects $\boldsymbol{m}_1$ and $\boldsymbol{m}_2$ differently. (b) Induced thermovoltage $\Delta V$ as a function of the angle difference. $\Delta_0$ is the bulk superconducting gap at zero temperature.
(c) Simulated angle difference for an external magnetic field $\mu_0H$ with four different RKKY couplings in units $\text{mJ/m²}$. (d) Thermovoltage as a function of the external magnetic field. For comparison, the inset shows the experimental results in Fig. \ref{Fig2}(a) for $\Delta T = 0.2$ K.}
\label{fig:numerical_model}
\end{center}
\end{figure}

The ferromagnets in Figure \ref{Fig1}(a) are treated as nonsuperconducting metallic reservoirs at temperatures $T$ and $T+\Delta T$ for the soft (MgO-Fe) and hard (Fe-Co) ferromagnets, respectively. The voltages in the reservoirs are $\pm \Delta V/2$, respectively. The voltages are chosen such that the electrical current $I$ through the system is zero. The interfaces to the superconductor are treated using spin-active tunneling interfaces with spin polarization $p$, tunneling conductance $G_0$, spin-mixing $G_{\varphi}$ and the average interface magnetization direction $\boldsymbol{m}$. The spin polarization filters incoming spins, while the spin-mixing generates a spin splitting in the superconductor. The average interface magnetizations lie in the plane of the interfaces, and the angle difference between the two interfaces is $\Delta \theta=\theta_2-\theta_1$. This angle difference depends on the external magnetic field, as explained in the previous section. The superconductor is treated as a Bardeen-Cooper-Schrieffer (BCS) \cite{bardeen_pr_57} superconductor of length $l$. The model is illustrated in Figure \ref{fig:numerical_model}(a). We assume that the system is diffusive, and we solve the Usadel equation \cite{usadel_prl_70} in the superconductor. This is done self-consistently for the superconducting order parameter. When the Usadel equation is solved, the current $I(\Delta V)$ is calculated. The thermovoltage is determined by solving $I(\Delta V)=0$.

\begin{figure}
    \centering
    \includegraphics[]{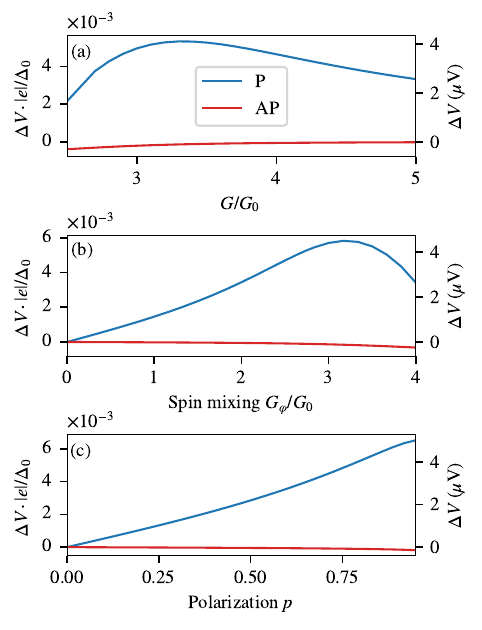}
    \caption{The thermovoltage dependence on (a) the ratio of the bulk conductance to the interface conductance $G/G_0$, (b) the spin mixing $G_{\varphi}$, and (c) the interface polarization $p$ in the parallel (P) and antiparallel (AP) configuration.}
    \label{fig:prameter dependence}
\end{figure}

In the simulations, we use the values $G_0/G=1/3.5$, $l=1.5\xi$ and $p=0.8$. Here, $G$ is the normal state bulk conductance and $\xi$ is the diffusive coherence length of the superconductor. The spin-mixing angle is typically taken to be a fitting parameter and is set to $G_{\varphi}/G_0=2.75$. The temperature is set to $T =0.057T_c^B$ and the temperature difference is $\Delta T=0.038T_c^B$, where $T_c^B$ is the bulk critical temperature of the superconductor. 
After solving the Usadel equation, we insert the numerical values $T_c^B=5.3$ K and $\xi=26$ nm \cite{GonzalezRuano2020, gonzalesruano_phd}, which yields $T = 0.3$ K, $\Delta T=0.2$ K, and $l=39$ nm. These are the values used to convert from dimensionless to numerical values in Figures \ref{fig:numerical_model} and \ref{fig:prameter dependence}. We also use the BCS relation $\Delta_0/k_BT_c^B = 1.76$ between the bulk gap $\Delta_0$ at zero temperature and the bulk critical temperature. The dependence of $\Delta V$ on $\Delta \theta$ is shown in Figure \ref{fig:numerical_model}(b). We note that with the chosen interface parameters, the quasiclassical simulations qualitatively reproduce the conductance curve in Figure \ref{Fig1}(c), although with a lower critical temperature of the superconducting film (not shown here).

A point requiring clarification is that while the temperature gradient and resulting thermovoltage are measured across the entire stack of materials in the experiment, the temperature gradient and thermovoltage are measured across the superconducting region in the theoretical modeling. However, this procedure is warranted for two reasons. As shown in \cite{supplemental}, the simulations for the local temperature profile reveal that the vast majority of the temperature drop occurs precisely across the Fe/MgO/V interfaces and the superconductor, and not in the ferromagnets themselves. Moreover, the voltage drop across the junction has its primary contribution coming from the interfaces between the ferromagnets and the superconductor due to the MgO-barrier providing a high resistance. This matches the assumptions made in the theoretical model.

We use micromagnetic simulations for four different RKKY couplings, as shown in Figure \ref{fig:numerical_model}(c), to model the thermoelectric response to an applied magnetic field. The thermovoltage as a function of the applied magnetic field is presented in Figure \ref{fig:numerical_model}(d), and does qualitatively match the experimental data in the inset. The difference in magnitude of the thermovoltage between the simulations and measurements may be attributed to the restriction in quasiclassical theory regarding how the reservoirs are treated. Normal and ferromagnetic reservoirs are described by the same Green's function due to the requirement of a weak spin polarization in the quasiclassical approximation, unlike the strongly polarized ferromagnet Fe used in the experiment, and this likely reduces the numerically simulated value for $\Delta V$. No such methodology restriction applies to the interfacial polarization in our treatment, however, which can model strongly polarized ferromagnetic interfaces ~\cite{eschrig_njp_15}. \\

\section{Discussion} 
Our experimental results demonstrate that the magnetic configuration of the superconducting spin-valve changes the 
thermoelectric signal by more than an order of magnitude when switching from a parallel to an antiparallel state. After achieving a qualitative agreement between the experimental results and simulations of both the TE and MR effects, we conclude that the thermoelectric effects are highly sensitive to the collective domain rotation within the ferromagnetic layers, as shown in \Cref{Fig3}(c). At $H=0$, due to the negative exchange interaction, the layers tend to align in an antiparallel configuration, resulting in the observed minima in resistance and TE signals. When an in-plane magnetic field is applied along one of the easy axes, the reorientation of the magnetization in the two ferromagnetic layers involves complex magnetic dynamics, including the formation of inner and edge domain walls, as illustrated in \Cref{Fig3}(c). 
Due to magnetocrystalline anisotropy, inner magnetic moments preferentially align towards the crystal’s easy axes, while the edge located spins tend to minimize their normal component to reduce the stray field related energy contribution.

In Figure~\ref{Fig3}, we also observe how the correlated rotation between the two ferromagnetic layers can induce an “overshoot” in the TE response (Figure~\ref{Fig2}(a)). The ”overshoot” is seen as a sharp dip that occurs in the angle misalignment between the average magnetizations in Figure~\ref{Fig3}(a).
This correlated rotation of two antiferromagnetically coupled Fe layers from points 2 to 4 as marked in Figure \ref{Fig3}(a) is explained by the corresponding magnetization maps in \ref{Fig3}(c). At point 3, one layer is mainly aligned along the x-axis, while the other aligns along the y-axis. At point 4, the alignment is interchanged. Following this transition, the layers continue to rotate towards a parallel alignment at high fields. This effect is observed in simulations and in the experiment only for AP $\rightarrow$ P  transition demonstrating the asymmetry between the P $\rightarrow$ AP and AP $\rightarrow$ P transitions.

The mechanism behind the thermoelectric effect was originally proposed in Refs.~\cite{machon_prl_13, ozaeta_prl_14}, and is described in detail in Ref. \cite{ouassou_prb_22}. The effect relies on two properties of the structure. First, the superconducting density of states must be spin split. Second, the interfaces must filter the spins such that the density of states available for the thermally excited electrons and holes is asymmetric around the Fermi energy. 

The thermoelectric effect is optimized when there is a large gap and spin splitting in the superconductor, and the interface polarization is high. For the parameter sets in Figure \ref{fig:numerical_model}, the gap value is approximately 75\% of the bulk value $\Delta_0$ for all angle configurations of the interfaces. If the interface conductance is good, corresponding to low $G/G_0$, the gap is suppressed due to the inverse proximity effect. This is seen in the P configuration in Figure \ref{fig:prameter dependence}(a).
As the interface conductance decreases, the gap value and thus the thermovoltage increases. By reducing the interface conductance further, the superconductor becomes more isolated from the reservoirs, and the transport across the superconductor is reduced. 
The spin mixing $G_{\varphi}$ induces the spin splitting in the superconductor, and the thermovoltage dependence on the spin mixing is shown in Figure \ref{fig:prameter dependence}(b). When $G_{\varphi}$ is zero, there is no spin splitting, while when $G_{\varphi}$ is high, the spin splitting is large and superconductivity is destroyed. The optimal $G/G_0$ and $G_{\varphi}$ values also depend on the length of the superconductor. A longer superconductor can withstand a higher interface conductance and spin mixing, thus increasing the thermovoltage. Moreover, sandwiching the superconductor in between the ferromagnets in the F/S/F structure increases the Seebeck coefficient compared to the S/F/F in Ref. \cite{gonzalez_prl_23} due to the increased spin splitting in the superconductor. Figure \ref{fig:prameter dependence}(c) shows that the thermovoltage increases with increasing interface polarization. This provides a way to optimize the thermoelectric effect.
As seen in Figure \ref{fig:prameter dependence}, the thermovoltage in the AP configuration is significantly smaller in magnitude than in the P configuration. 
This is because the average spin splitting in the AP configuration is smaller than in the P configuration.  

Finally, we note that the large Seebeck coefficient in the P configuration remains large and more than an order of magnitude greater than in the AP configuration as the base temperature is increased, meaning that the effect is robust against an increase in temperature \cite{supplemental}. These theoretical predictions will be tested and verified in forthcoming experiments currently in preparation.

\section{Conclusions} 
We report the experimental observation of
a large superconducting thermoelectric spin-valve effect which meets a number of key criteria required for cryogenic thermoelectric applications: (i) we use commercially
available materials with high spin polarization which are easily
grown; (ii) a considerable thermoelectric effect, with a Seebeck coefficient exceeding 100 $\mu$V/K, (iii) a large change
in the thermoelectric response – greater than a factor of 10 –
when switching the device from AP to P alignment. Micromagnetic simulations and 
theoretical modeling
based on the non-equilibrium Keldysh-Usadel Green's function
formalism are consistent with both the thermoelectric and magnetoresistance results. As real cryogenic quantum systems are expected to operate under thermal gradients, integrating low-temperature Peltier elements into quantum computers could significantly enhance their performance. One promising future application of ferromagnet/superconductor/ferromagnet-based devices is the recovery of heat inevitably generated during quantum error correction \cite{bilokur2024Arxiv}. This is particularly relevant since most current quantum computers use qubits based on superconducting hybrid structures.

\section{Methods}

A more detailed explanation of the matters treated in this section is provided in the Supplemental Material~\cite{supplemental}.\\

\paragraph{Sample growth and experimental procedure:}
The F/S/F junctions studied in this work were grown by molecular beam epitaxy (MBE) and litographed into $20\times20~\mu\text{m}^2$ lateral size samples. Both MR and TE response measurements highlighted in Figure~\ref{Fig2} were performed at temperatures of 0.3~K and 11~K since these temperatures can be sustained in the cryogenic system (Janis He$^3$ cryostat) without the use of external heaters, minimizing extrinsic thermal gradients. The TE voltage $\Delta{V}$ results from subtracting the background voltage signal from the voltage measured under heating at zero current bias. Different temperature gradients are induced by controlling the power dissipated by the LED by tuning its voltage bias $V_{LED}$. In MR measurements performed at thermal equilibrium, voltage is measured at bias currents of $I=\pm10$~nA and the resistance is calculated as $R=V/(2|I|)$. This current bias yields a voltage of $\sim5$~mV.\\

\paragraph{Modeling of the temperature profile:}
In order to estimate the total temperature difference $\Delta{T}$ across the junction when heated under different $V_{LED}$, we solve the heat diffusion equation in a 1D model of our junctions. We impose an inward heat flux $q$ on one of the edges to account for the LED heating. The value of $q$ is estimated by taking into account the power dissipated by the LED ($I_{LED}\times{V}_{LED}$) and its directionality through the radiation pattern provided by the manufacturer, the distance between the LED and the sample, and the contact area of the junctions that can absorb heat. The other edge of the system is fixed at a given temperature. We use the calculated values of $\Delta{T}$ to estimate the Seebeck coefficient under different magnetic fields, base temperatures, and temperature gradients.\\

\paragraph{Micromagnetic simulations:}
The relative angle between the ferromagnetic layers is calculated using micromagnetic simulations performed in MuMax$^3$ \cite{Mumax2014}. We simulate a 3D model of the magnetic layers our junctions with matching vertical dimensions. A phenomenological antiferromagnetic coupling is set between the two Fe layers of the system to reproduce the behavior under applied in-plane magnetic fields inferred from the MR measurements in Figure~\ref{Fig2}(c). Superficial and bulk defects are implemented by lowering the saturation magnetization and adding disorder to the local magnetocrystalline anisotropy, respectively. The calculated MR displayed in Figure~\ref{Fig3}(b) is obtained by using the Slonczewski formula~\cite{PhysRevB.39.6995} with a sign change to account for the negative MR induced by the presence of the V layers in between the two ferromagnetic electrodes.\\

\paragraph{Thermoelectricity via non-equilibirum quasiclassical theory:}
We use non-equilibrium Keldysh-Usadel Green's function theory \cite{bergeret_rmp_18, theta_parameterization, usadel_prl_70} to model the thermoelectric effect in the F/S/F structure numerically. The ferromagnets are treated as non-superconducting metallic reservoirs with spin-active interfaces. We calculate the quasiclassical Green's function in the superconducting layer numerically by solving the Usadel equation self-consistently for the superconducting order parameter, and find the voltage $\Delta V$ such that the charge current through the system disappears. The thermovoltage $\Delta V$ depends on the external magnetic field through the relative angle of the average interface magnetizations. 

\acknowledgments
\textit{Acknowledgments.} The work in Madrid was supported by Spanish Ministry of Science and Innovation (PID2021-124585NB-C32, TED2021-130196B-C22 and PID2024-155399NB-I00). F.G.A. also acknowledges financial support from the Spanish Ministry of Science and Innovation through the Mar\'ia de Maeztu Programme for Units of Excellence in R\&D (CEX2023-001316-M) and TEC-2024/TEC-380 (Mag4TIC-CM) project. The work in Trondheim was supported by the Research Council of Norway through Grant No. 323766 and its Centres
of Excellence funding scheme Grant No. 262633 “QuSpin.” Support from
Sigma2 - the National Infrastructure for High Performance
Computing and Data Storage in Norway, project NN9577K, is acknowledged.

\bigskip

$^{\dag}$P.T. and J.B.T. contributed equally to the manuscript.

\bibliography{references}

\end{document}


\title{Supplemental Material for \textit{Large tunable thermoelectric effects in superconducting \\spin valves with commercially available materials}}

\author{Pablo Tuero}
\affiliation{Departamento F\'isica de la Materia Condensada C-III, Universidad Aut\'onoma de Madrid, Madrid 28049, Spain}
\affiliation{These authors contributed equally to the manuscript}

\author{Johanne Bratland  Tjernshaugen }
\affiliation{Center for Quantum Spintronics, Department of Physics, Norwegian University of Science and Technology, NO-7491 Trondheim, Norway}
\affiliation{These authors contributed equally to the manuscript}

\author{Carlos Sanchez}
\affiliation{Departamento F\'isica de la Materia Condensada C-III, Universidad Aut\'onoma de Madrid, Madrid 28049, Spain}

\author{César Gonzalez-Ruano}
\affiliation{Departamento F\'isica de la Materia Condensada C-III, Universidad Aut\'onoma de Madrid, Madrid 28049, Spain}

\author{Yuan Lu}
\affiliation{Université de Lorraine, CNRS, IJL, F-54000 Nancy, France}

\author{Jacob Linder}
\email[e-mail: ]{jacob.linder@ntnu.no}
\affiliation{Center for Quantum Spintronics, Department of Physics, Norwegian University of Science and Technology, NO-7491 Trondheim, Norway}

\author{Farkhad G. Aliev}
\email[e-mail: ]{farkhad.aliev@uam.es}
\affiliation{Departamento F\'isica de la Materia Condensada C-III, Instituto Nicol\'as Cabrera (INC) and  Condensed Matter Physics Institute (IFIMAC), Universidad Aut\'onoma de Madrid, Madrid 28049, Spain}

\date{\today}



\begin{abstract}

\end{abstract}


\maketitle



\makeatletter 
\renewcommand{\thefigure}{S\@arabic\c@figure}
\makeatother

\onecolumngrid
\noindent

\section{Sample growth and experimental procedure}

\begin{wrapfigure}{l}{0.35\textwidth}
\begin{center}
\includegraphics[width=1\linewidth]{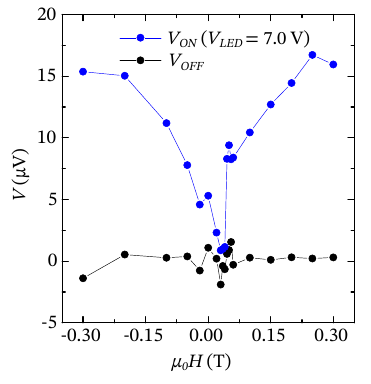}
\caption{Magnetic field dependence of the voltage without LED
heating ($V_{OFF}$) and the voltage measured under LED heating ($V_{ON}$) with $V_{LED}=7.0$~V which corresponds with the $T = 0.3$~K and $\Delta{T}=0.13$~K data of Figure 2(a,b) of the main text.}
\label{fig_S1}
\end{center}
\end{wrapfigure}

The F/S/F junctions under study are grown by molecular beam epitaxy (MBE) with their crystalline quality controlled by in-situ RHEED measurements. The structures are lithographed into 20x20 $\mu$m$^{2}$ lateral size samples. A square shape complements the cubic anisotropy, leading to a more symmetric energy landscape and more predictable switching behavior compared to a non-rectangular sample shape. A more detailed description of the growth process can be found in Ref. \cite{Tiusan_2007}.

\vspace{4pt}

In order to measure zero the field superconducting critical temperature through the temperature dependence of the vertical conductance of the FSF stack (Figure 1(a) of the main text without heat gradients imposed by the Light Emitting Diode (LED), local sample heaters or cryopump heaters, we first maintain (for about 15 min) the sample at $T\sim20$~K with a local sample heater. Then we let the sample slowly cool down (during a few hours) to $T\sim2$~K by continuously pumping the 1~K (He4) port only. 
However, to reach lower temperatures, one needs to condense and cycle He3 through the 1~K port cooling system by means of a cryopump. This cryopump requires the use of a second separate heater to control its temperature. Thus, technically, we cannot measure resistance versus temperature in the $0.3-2$~K range in the same thermodynamic quasi-equilibrium as we manage to reach when cooling from well above to below the superconducting $T_c$, as we cannot simultaneously use the thermometer to measure the sample’s temperature and to control the cooling process by cryopump with its heater and thermometer.

\vspace{4pt}

Magnetoresistance (MR) and thermoelectric (TE) response measurements are performed inside a Janis He$^{3}$ cryostat at temperatures of 0.3~K and 11~K. At these temperatures, the external heater needs not be used to stabilize the base temperature and thus the temperature gradients from the external environment are minimized. Therefore, the conditions closest to thermal equilibrium have been used to measure the resistance of the sample and the background thermoelectric voltage before creating heat gradients with the LED. For both kinds of measurements, the voltage drop through the sample $V$ is measured at a given applied current $I$ with a four-point probe set up. The magnetic field $H$ is applied along the easy axis of the samples. The TE response experiments are performed with the use of a commercial LED, in a similar way as explained in Ref. \cite{gonzalez_prl_23}. The voltage $V_{LED}$ is supplied by a SRS-DC205 voltage source to a LED placed above the samples. By controlling $V_{LED}$ one can tune the power that the diode dissipates, creating a temperature gradient $\Delta T$ across the junctions. A TE measurement at a given $V_{LED}$ is carried out as follows. First, forcing $I=0$ through the sample at the base temperature, the voltage across the sample ($V_{OFF}$) is measured while $V_{LED}=0$. Then, the target $V_{LED}$ is applied and after two seconds the voltage across the junctions is again measured ($V_{ON}$). The TE voltage is then $\Delta{V}(V_{LED})=V_{ON}-V_{OFF}$. An example of the field dependence of $V_{ON}$ and $V_{OFF}$ is provided in Figure~\ref{fig_S1}. The background signal $V_{OFF}$ appears possibly due to the unavoidable heat gradients between the He3 cold bath at 0.3 K and the electric manganin cables that, although thermally poor conductors, carry heat from warmer heat sources. The estimation of $\Delta T$ as a function of $V_{LED}$ in order to compute $\Delta{V}(\Delta T)$ is discussed in the following section.

\vspace{4pt}

To evaluate the MR, the voltage is measured at currents $I=\pm10$~nA, so that the resistance is calculated as $R=V/(2|I|)$. Resistance is measured at a given magnetic field before measuring the thermoelectric response, that is, before applying any $V_{LED}$. After a thermoelectric response measurement the next magnetic field is set and we let the system thermalize back to 0.3 K or 11 K for several minutes before the next magnetoresistance measurement is taken.

\section{Modeling of the temperature profile}

\begin{figure}[H]
\begin{center}
\includegraphics[width=0.9\linewidth]{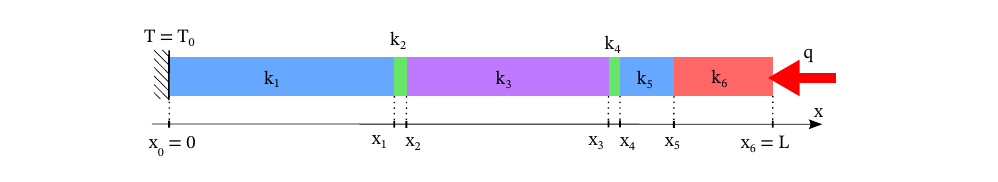}
\caption{Sketch of the 1D model used to solve the heat diffusion equation. Positions $x_{i}$ indicate the boundaries of the different material regions and $k_{i}$ is the thermal conductivity of the $i^{th}$ material region. The system is fixed at a constant temperature $T_{0}$ on the boundary and a constant inward heat flux $q$ is imposed on the right boundary.}
\label{fig_S2}
\end{center}
\end{figure}

In order to calculate the temperature difference across the junction under different values of $V_{LED}$ we analytically solve the stationary heat equation [Eq.\eqref{Heat_Eq}] in a 1D system that approximates the vertical structure of the our samples, sketched in Figure~\ref{fig_S2}. We impose an inward heat flux $q$ through the top surface (right edge in the 1D model) and assume that the lateral faces of the stack are thermally isolated (the samples are in a vacuum). Finally, we fix the temperature value in the bottom surface (left edge in the 1D model) to the temperature of the cryostat's thermal bath, with which it is in contact. In this approximated 1D system, the problem to solve is then
\begin{equation}
    \frac{d}{dx}\left\{k\left(x\right)\frac{d}{dx}\Bigl[T\left(x\right)\Bigr]\right\}=0,
    \label{Heat_Eq}
\end{equation}
with the space-dependent thermal conductivity $k(x) = k_{i}\mid x\in[ x_{i-1}, x_{i})$. The index $i\in\left\{1,...,6\right\}$ indicates the $i^{th}$ material region from left to right, as sketched in Figure~\ref{fig_S2}.
The boundary conditions are
\begin{equation}
    \text{(a)}~~T(x=0) = T_0~;\quad\quad\text{(b)}~~\left\{-k\left(x\right)\frac{d}{dx}\Bigl[T\left(x\right)\Bigr]\right\}_{x = L} = -q.
    \label{Boundaries}
\end{equation}
Inside a given material region, we have $k\frac{d^2}{dx^2}\Bigl[T\left(x\right)\Bigr]=0$, so the solution will have the form $T(x)  = x\cdot m_{i}+n_{i} \mid x\in[ x_{i-1}, x_{i})$. From Eq. (\ref{Boundaries}a) it now follows that $n_{1} = T_{0}$, and from Eq. (\ref{Boundaries}b) that $m_{6}=q/k_{6}$. If we now impose that the solution must be continuous:
\begin{equation}
    x_{i} m_{i}+n_{i}=x_{i} m_{i+1}+n_{i+1},
    \label{Continuity}
\end{equation}
and we assure that Eq. \eqref{Heat_Eq} is fulfilled in the material region boundaries:
\begin{equation}
    \left\{-k_{i}\frac{d}{dx}\Bigl[T_{i}\left(x\right)\Bigr]\right\}_{x = x_{i}^{-}} = \left\{-k_{i+1}\frac{d}{dx}\Bigl[T_{i+1}\left(x\right)\Bigr]\right\}_{x = x_{i}^{+}} \longrightarrow k_{i}m_{i} = k_{i+1}m_{i+1},
    \label{MaterialBoundaries}
\end{equation}
it follows from Eq.\eqref{MaterialBoundaries} that $k_{i}m_{i} = k_{6}m_{6}$. Substitute the value found for $m_{6}$:
\begin{equation}
m_{i} = q/k_{i}.
\label{slopes}
\end{equation}

Knowing the values of $m_{i}$ and $n_{1}$ we can now write an expression for $n_{i}$ by substituting regressively in Eq.\eqref{Continuity}:
\begin{equation}
    n_{i} = T_{0}+\sum_{j=2}^{i}x_{j-1}\left(m_{j-1}-m_{j}\right) = T_{0}+\sum_{j=2}^{i}x_{j-1}\left(\frac{q}{k_{j-1}}-\frac{q}{k_{j}}\right).
\end{equation}
Thus, the temperature difference between the top and bottom of the system, which is defined as $\Delta{T}=T(x=L)-T(x=0)$, as a function of the heat flux $q$ absorbed in the $x=L$ boundary is:
\begin{equation}
    \Delta{T}(q) = q\Biggl\{\frac{L}{k_{6}}+\sum_{i=2}^{6}x_{i-1}\left(\frac{1}{k_{i-1}}-\frac{1}{k_{i}}\right)\Biggr\}=qL\sum_{i=1}^{6}\frac{x_{i}-x_{i-1}}{k_{i}}\cdot\frac{1}{L}=\frac{qL}{k_{eq}}
    \label{DeltaT}
\end{equation}
The value of $k_{eq}$ depends only on the geometry and materials of the system, and it is the geometric average of the values of $k_{i}$ weighted by the length of their respective material region. We use tabulated values for the different material parameters in the low temperature regime \cite{Anoop2021,Gardner_1981,chakalskii_1978,Balcerek1996} given in Table.\ref{table:heat}. Importantly, the top MgO barrier is expected to be $\sim4$ times more transparent than the bottom one \cite{Martinez2018}, so we have proportionally tuned its thermal conductivity. Computing the value of $k_{eq}$ yields $k_{eq}=11.57~\text{W}/\text{(m}\cdot\text{K)}$.

We believe this is an underestimated value since it does not take into account the small thickness of the MgO films present in our samples nor the symmetry dependent tunneling effects where the $k_{||}=0$ "hotspots" in momentum space are expected to contribute to electron and heat transport through MgO from Fe \cite{Butler2001}. We note that the thermal conductivity of superconducting vanadium used in the calculations could also be underestimated as the experimentally observed finite value of zero bias conductance
(Figure 1(b) in the main text) could be associated with some additional thermal conductance.

\begin{table}
\begin{center}
\begin{tabular}{ |>{\centering\arraybackslash}m{2.2cm}|>{\centering\arraybackslash}m{2.2cm}||>{\centering\arraybackslash}m{2.2cm}|>{\centering\arraybackslash}m{2.2cm}|  }
 \hline
 \multicolumn{4}{|c|}{Constants of the 1D model} \\
 \hline
 $k_{1}$ & $k_{Fe}$   & $x_{1}$ & 45 nm\\
 $k_{2}$ & $k_{MgO}$  & $x_{2}-x_{1}$ & 2.3 nm\\
 $k_{3}$ & $k_{V}$    & $x_{3}-x_{2}$ & 40 nm\\
 $k_{4}$ & $4k_{MgO}$ & $x_{4}-x_{3}$ & 2.3 nm\\
 $k_{5}$ & $k_{Fe}$   & $x_{5}-x_{4}$ & 10 nm\\
 $k_{6}$ & $k_{Co}$   & $x_{6}-x_{5}$ & 20 nm\\
 \hline
 \multicolumn{4}{|c|}{\multirow{2}{*}{$k_{Fe}=70~$\cite{Anoop2021};$\quad k_{MgO}=0.1~$\cite{Gardner_1981};$\quad k_{V}=10~$\cite{chakalskii_1978};$\quad k_{Co}=100~$\cite{Balcerek1996};$\quad
 \left[\text{W/(m}\cdot\text{K)}\right]$}} \\
 \multicolumn{4}{|c|}{\multirow{2}{*}{ }}\\
 \hline
\end{tabular}
\caption{Thermal conductivities and thicknesses of the different material regions used to reproduce the vertical structure of our samples in the 1D model in order to calculate the temperature profiles.}
\label{table:heat}
\end{center}
\end{table}

Now we calculate the value of $q$ for the different values of applied $V_{LED}$. To do that, the room temperature $I-V$ curve of the diode provided by the manufacturer (model LUXEON~3030~2D) is re-scaled to match the working range of the LED at low temperatures, knowing that heating started when $V_{LED}> 6.5$~V and that $I_{LED}= 100$~mA when $V_{LED}=7.1$~V. Then, for a given $V_{LED}$ the total power dissipated by the device is $Q=I_{LED}\times{V_{LED}}$. In order to know the power flux that reaches the area of the sample, we take into account the radiation pattern characteristic $f(\theta)$ provided by the LED manufacturer. $\theta$ is the polar angle defined so that $\theta=0$ for the direction perpendicular to the LED surface and facing the samples. As one may expect, the LED does not radiate energy isotropically. Instead, it has a nonuniform distribution centered in the direction in which the LED is facing and that is effectively zero on its backside. Since the LED must dissipate a total power $Q$, we renormalize the curve provided by the manufacturer so that:
\begin{equation}
    2\pi{R^2}\int_{0}^{\pi}f(\theta)sin(\theta)d\theta=1
    \label{LED_normalization}
\end{equation}

\begin{figure}[h]
\begin{center}
\includegraphics[width=1\linewidth]{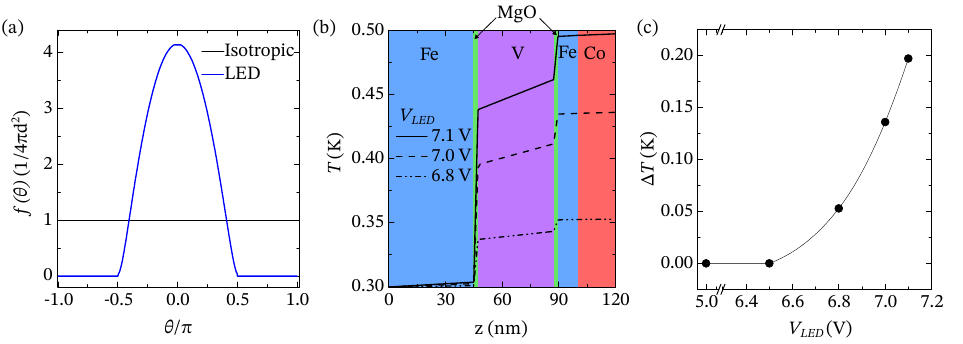}
\caption{(a)~Radiation pattern characteristics for an isotropic emitter and for the LED used in our experiments. The curve for the LED was obtained by normalizing the curve provided by the manufacturer with Eq.\ref{LED_normalization}. (b)~Temperature profiles for the different $V_{LED}$ relevant to our experiments. The colors indicate the layers and materials of the heterostructures. For the values of $V_{LED}$ of 7.1, 7.0 and 6.8 V the applied current $I_{LED}$ is 100, 65 and 30 mA, respectively. (c)~The estimated temperature difference between the top and bottom of the samples as a function of the applied $V_{LED}$.}
\label{fig_S3}
\end{center}
\end{figure}
This angular distribution is shown together with the characteristic curve of an isotropic pattern in Figure~\ref{fig_S3}(a). With this expression, we find that $Qf(\theta=0)$ is the power per unit area that reaches the sample. We also assume that the LED is heating the area of the gold contact atop the samples (1~mm$^2$), and that all the heating power that reaches the contact is transmitted into the upper layer of the sample. With all this information, we get an expression for the inward heat flux $q$ that we need for our calculations:
\begin{equation}
    q = {I_{LED}V_{LED}}{f(\theta=0)}\frac{A_\text{contact}}{A_\text{sample}}
    \label{q_expression}
\end{equation}
where $A_\text{contact}$ and $A_\text{sample}$ are the section of the 
gold contacts and the sample under study, respectively.

\vspace{4pt}

\begin{figure}[h]
\begin{center}
\includegraphics[width=1\linewidth]{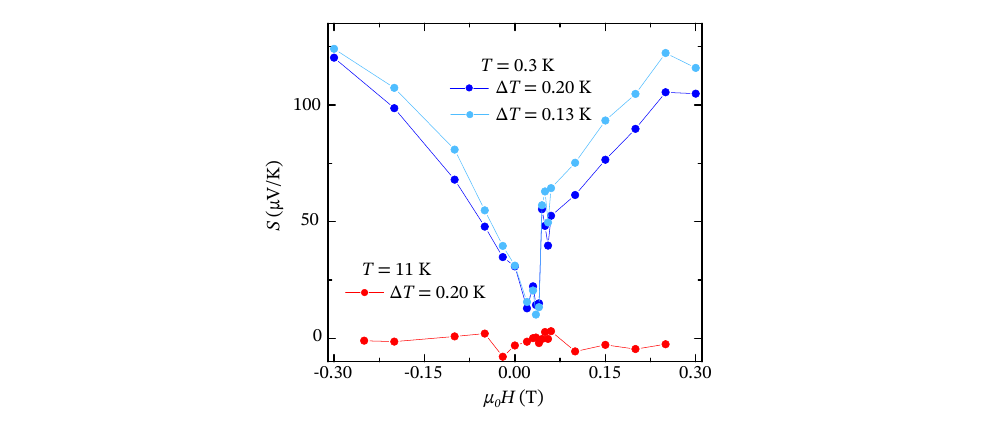}
\caption{Seebeck coefficient as a function of the applied in-plane magnetic field at two different base temperatures: deep in the superconducting regime ($T=0.3~\text{K}$) and well above the superconducting critical temperature ($T=11~\text{K}$). Note that the measured superconducting critical temperature is $T_{c}=4.7~\text{K}$ (see Figure 1c of the main text). The Seebeck coefficient was calculated using the measured data for the thermoelectric voltage shown in Figure 2a of the main text and the estimations of $\Delta{T}$ explained in this section.}
\label{fig_S4}
\end{center}
\end{figure}

Finally, in Figure~\ref{fig_S3}(b,c) we show the calculated temperature profile across the junction for different $V_{LED}$ and the temperature difference $\Delta{T}$ between the upper and lower layers of the sample, respectively. Using these values for $\Delta{T}$ and the measured $\Delta{V}$ shown in Figure 2(a) of the main text, one can get an estimation of the Seebeck coefficient $S$ of the junction and its variation with temperature and magnetic field, which is shown is Figure~\ref{fig_S4}. We obtain a maximum Seebeck coefficient of $\sim125$~$\mu$V/K in the P state and a minimum of $\sim10$~$\mu$V/K in the AP state when $T=0.3$~K. When $T=11$~K, above vanadium superconducting critical temperature, we obtain values for the Seebeck coefficient very close to zero ($ \lesssim3~\mu$V/K)  regardless of the magnetic configuration in the junction.

\section{Micromagnetic simulations methods}
\begin{figure}[h]
\begin{center}
\includegraphics[width=1\linewidth]{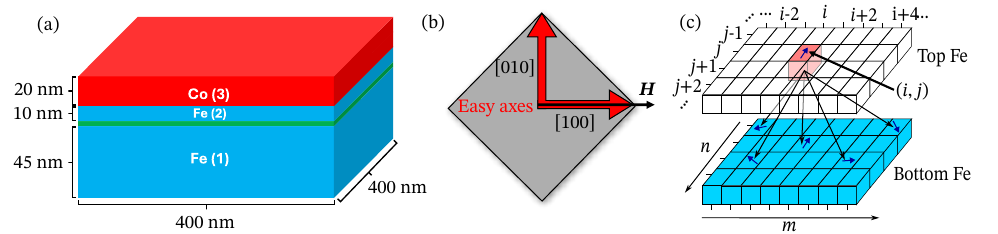}
\caption{
(a) Sketch of the ferromagnetic junctions used in micromagnetic simulations. (b) Indications of the cubic magnetocrystalline anisotropy of the system and direction of the in-plane magnetic field applied. (c) Sketch describing Rnd.Diff. method of magnetoresistance calculation.}
\label{fig:simulation}
\end{center}
\end{figure}
The micromagnetic simulations are perfomed in MuMax$^3$, 
and the simulated system is presented in \Cref{fig:simulation}(a,b). It consists of a $400\times400\times75$~nm cuboid with a discretitation cell size of $2.34\times2.34\times1.67$~nm, smaller than exchange lengths for Fe and Co. The system is divided into three separated regions with different magnetic parameters and couplings between them to mimic the experimental setup in Fig. 1(a). All these regions have a base of $400\times400$~nm. 
To simulate a more realistic system, the presence of defects in the crystalline structure of the layers is taken into account. We make a distinction between two types of defects. First, superficial defects are only implemented in the first and last cell layers in Regions 1 and 3, and present a lower saturation magnetization than the normal material ($M_S^{\text{defect}}=0.6M_S$). Second, bulk defects are introduced by adding disorder to the local magnetocrystalline anisotropy. To do that, the cells modelling bulk defects have a randomly distributed anisotropy direction and a change in the 1st order cubic anisotropy constant, in the range of $(1\pm0.15)K_{1}$. In this paper we use $K_1= 4.8\cdot 10^4 $ J/m$^3$ \cite{Graham1958}. In the same way, the principal easy axes of magnetization are set to [100] and [010], as indicated in \Cref{fig:simulation}(b).

To calculate the magnetoresistance we follow a simplified model~\cite{PhysRevB.39.6995} to approximate the magnetic state dependent conductance $G$ for our junctions with the following expression:
\begin{equation}
   G=G_0(1 - p^2\cos\Delta\theta).
   \label{Eq1}
\end{equation}
Here, $G_0$ is the low bias conductance of the junctions in a perpendicular state, $p$ is the effective spin polarization, and $\Delta\theta$ is the angle between the magnetizations of the ferromagnets. 

For these calculations, we only take into consideration the interfacial layers of cells between regions 1 and 2. We consider two possible ways of calculating the total resistance G$^{-1}$ of the junction. One (Av.Angle) method consists of direct application of Eq.\eqref{Eq1} using the averaged magnetization angles between the regions. The second (Rnd.Diff.) considers every cell $(i,j)$ of region 1 and calculates, using eq. \eqref{Eq1}, the conductance to a randomly chosen cell $(n,m)$ in region 2 as sketched in \Cref{fig:simulation}(c). Finally, we sum the contribution from all pairs in parallel to obtain the total conductance and resistance for the junction. We have verified that the resulting conductance does not change when the top and bottom Fe regions (1 and 2) are interchanged. Several repetitions of the Rnd.Diff. MR calculation method have been also carried out to verify robustness of the Rnd.Diff. method.

\section{Thermoelectricity via non-equilibrium quasiclassical theory}

We model the thermoelectric effect in the system shown in Figure 4(a) in the main text using the self-consistent non-equilibrium Keldysh-Usadel Green's function theory \cite{bergeret_rmp_18, theta_parameterization}. This formalism is known to compare well with experiments in mesoscopic superconducting hybrid systems. The $8\times 8$ Green's function $\check{G}$ in Keldysh$\otimes$Nambu$\otimes$ spin space is defined as
\begin{equation}
    \check{G}=\begin{pmatrix}
        \hat{G}^R & \hat{G}^K \\ 0 & \hat{G}^A
    \end{pmatrix}.
\end{equation}
Here, the retarded Green's function $\hat{G}^R$, the advanced Green's function $\hat{G}^A$ and the Keldysh Green's function $\hat{G}^K$ are defined in terms of the spinor  
\begin{equation}
    \psi(\boldsymbol{r},t) = \begin{pmatrix}
        \psi_{\uparrow} (\boldsymbol{r},t) & \psi_{\downarrow}(\boldsymbol{r},t) & \psi_{\uparrow}^{\dagger}(\boldsymbol{r},t) & \psi_{\downarrow}^{\dagger}(\boldsymbol{r},t)
    \end{pmatrix}^T,
\end{equation}
as
\begin{align*}
    \hat{G}^R(\boldsymbol{r},t,\boldsymbol{r}',t') &= -i\theta(t-t')\hat{\rho}_4\langle \{ \psi(\boldsymbol{r},t), \psi^{\dagger}(\boldsymbol{r}',t') \} \rangle, \\   \hat{G}^A(\boldsymbol{r},t,\boldsymbol{r}',t') &= +i\theta(t'-t)\hat{\rho}_4\langle \{ \psi(\boldsymbol{r},t), \psi^{\dagger}(\boldsymbol{r}',t') \} \rangle, \\ \hat{G}^K(\boldsymbol{r},t,\boldsymbol{r}',t') &= -i\hat{\rho}_4\langle [ \psi(\boldsymbol{r},t), \psi^{\dagger}(\boldsymbol{r}',t') ] \rangle.
\end{align*}
Here, $\psi_{\sigma}^{\dagger}$ and $\psi_{\sigma}$ are the electronic creation and annihilation operators, respectively. The matrix $\hat{\rho}_4$ is a part of a set of matrices $\{\hat{\rho}_n\}$ that span the block-diagonal Nambu$\otimes$spin space,
\begin{align}
    \hat{\rho}_0 =\begin{pmatrix}
        \sigma_0 & 0 \\ 0 & \sigma_0
    \end{pmatrix} \quad 
    \hat{\rho}_1 =\begin{pmatrix}
        \sigma_1 & 0 \\ 0 & \sigma_1
    \end{pmatrix} \quad 
    \hat{\rho}_2 = \begin{pmatrix}
        \sigma_2 & 0 \\ 0 & \sigma_2 
    \end{pmatrix}\quad 
    \hat{\rho}_3 =\begin{pmatrix}
        \sigma_3 & 0 \\ 0 & \sigma_3
    \end{pmatrix} \\
     \hat{\rho}_4 =\begin{pmatrix}
        \sigma_0 & 0 \\ 0 & -\sigma_0
    \end{pmatrix} \quad 
    \hat{\rho}_5 =\begin{pmatrix}
        \sigma_1 & 0 \\ 0 & -\sigma_1
    \end{pmatrix} \quad 
    \hat{\rho}_6 = \begin{pmatrix}
        \sigma_2 & 0 \\ 0 & -\sigma_2 
    \end{pmatrix}\quad 
    \hat{\rho}_7 =\begin{pmatrix}
        \sigma_3 & 0 \\ 0 & -\sigma_3
    \end{pmatrix}. 
\end{align}
The Pauli matrices are
\begin{align}
    \sigma_0=\begin{pmatrix} 1&0\\0&1\end{pmatrix} \quad \sigma_1 = \begin{pmatrix}
        0&1\\1&0
    \end{pmatrix} \quad \sigma_2=\begin{pmatrix}
        0&-i\\i&0
    \end{pmatrix} \quad \sigma_3=\begin{pmatrix}
        1&0\\0&-1
    \end{pmatrix}.
\end{align}

The exact Green's function $\check{G}$ can in principle be determined from the Gorkov equation \cite{gorkov_energy_1958}, but in practice this is too hard and one relies on simplifying approximations for all but the simplest systems. In the quasiclassical limit, all energy scales in the system are much smaller than the Fermi energy and $\check{G}$ is strongly peaked near the Fermi momentum $\boldsymbol{p}_F$~\cite{belzig_micro_99, chandrasekhar_superconductivity_2008}. The quasiclassical Green's function $\check{\Gamma}$ is found by constricting the Green's function to the Fermi surface,
\begin{equation}
    \check{\Gamma}\left(\varepsilon, \frac{\boldsymbol{p}_F}{|\boldsymbol{p}_F|}\right) = \frac{i}{\pi} \int_{-\omega_c}^{\omega_c} \check{G}(\varepsilon, \boldsymbol{p}) \text{d}\epsilon_{\boldsymbol{p}},  
\end{equation}
where $\epsilon_{\boldsymbol{p}}= {(|\boldsymbol{p}|^2-|\boldsymbol{p}_F|^2)}/{2m}$, and $\varepsilon$ is measured relative to the Fermi energy. The quasiclassical Green's function obeys the Eilenberger equation~\cite{eilenberger_zphys_68}. For diffusive systems where the elastic mean free path is much smaller than any other length scale in the system except the Fermi wavelength, the Eilenberger equation can be reduced to a diffusion-like equation. This is known as the Usadel equation \cite{usadel_prl_70},
\begin{equation}
    {\frac{\partial}{\partial (x/\xi)}\left(\check{g} \frac{\partial\check{g}}{\partial (x/\xi)}\right) }= -i[\varepsilon\hat{\rho}_4+\hat{\Delta},\check{g}]/\Delta_0.
\end{equation}
Here, $\check{g}$ is the isotropic, quasiclassical Green's function, which we term the Green's function. The superconducting coherence length is $\xi$, and $\hat{\Delta} = \text{antidiag}(\Delta, -\Delta, \Delta^*, -\Delta^*)$ is a self-energy describing superconductivity in the system. The superconducting order parameter is defined as $\Delta(\boldsymbol{r}) = -\lambda \langle \psi_{\downarrow}(\boldsymbol{r}) \psi_{\uparrow}(\boldsymbol{r}) \rangle$, and the bulk superconducting gap at zero temperature is $\Delta_0$. Inelastic scattering in the Usadel equation is modeled using the Dynes approximation $\varepsilon \rightarrow \varepsilon + i\delta$.

On matrix form, the Green's function is written as \begin{equation}
    \check{g} = \begin{pmatrix}
        \hat{g}^R & \hat{g}^K \\ 0 & \hat{g}^A
    \end{pmatrix}.
\end{equation}
The retarded and advanced Green's functions are related by 
 $\hat{g}^A = - \hat{\rho}_4 (\hat{g}^R)^{\dagger}\hat{\rho}_4$. The Keldysh Green's function is related to those by
\begin{align}
\hat{g}^K = \hat{g}^R\hat{h}- \hat{h}\hat{g}^A,
\end{align}
where $\hat{h} = h_n \hat{\rho}_n$ is the distribution function. Therefore, it is sufficient to determine the retarded Green's function and the distribution function when solving the Usadel equation. The retarded Green's function was Riccati parametrized \cite{schopohl_quasiparticle_1995,konstandin_superconducting_2005} and the distribution function was parametrized with a trace parameterization \cite{bergeret_rmp_18, theta_parameterization, ouassou_prb_18} to solve the Usadel equation. 

The distribution functions in the reservoirs determine the temperature and voltage gradients across the superconductor. The non-equilibrium modes $h_n$ in the distribution function describe the relation between the occupation of electron and hole states with different spins. Holes are defined as missing electrons, and the hole has opposite spin to the missing electron. The distribution function can be rewritten in terms of the electron and hole occupation probabilities $f_{e,\sigma}$ and $f_{h,\sigma}$ as \cite{distributionfnc-form}
\begin{equation}\label{eq: disfnc in terms of occupation}
    \hat{h} = \hat{1} - 2\cdot \text{diag}(f_{e,\uparrow}, f_{e,\downarrow},f_{h,\downarrow}, f_{h,\uparrow}).
\end{equation}
In equilibrium, the electron and hole occupation probabilities are given by the Fermi-Dirac distribution $f(\varepsilon)$. In the presence of an electrical voltage $V$, the distribution function becomes 
\begin{equation}\label{eq:disfnc for a standard voltage}
    \hat{h}=\text{diag}(t_+, t_+, t_-, t_-),
\end{equation}
where 
\begin{equation}
  t_{\pm} = \text{tanh}\left( \frac{1.76}{T/T_c} \cdot \frac{\varepsilon/\Delta_0 \pm |e|V/\Delta_0}{2} \right).
\end{equation}
Here, we have used that the bulk gap and the critical temperature are related through the BCS relation $\Delta_0/k_BT_c = 1.76$.
A temperature gradient in a superconductor interfaced with two reservoirs is modeled by setting the temperature in the reservoirs to different values. Similarly, a voltage bias $\Delta V$ is modeled by setting the voltage to $+\Delta V/2$ in one reservoir and $-\Delta V/2$ in the other reservoir.  

The superconductor is coupled to the normal reservoirs with spin-active boundary conditions~\cite{eschrig_njp_15, ouassou_prb_18}:
\begin{equation}
    \check{g}\frac{\partial}{\partial(x/\xi)}\check{g} = \frac{\pm G_0/G}{2 l/\xi} \left[\check{g}, \underline{\check{g}} + \frac{G_1}{G_0} \check{m}\underline{\check{g}}\check{m} + \frac{G_{MR}}{G_0}\{\underline{\check{g}}, \check{m}\} -i \frac{G_{\varphi}}{G_0}\check{m}' \right].
\end{equation}
Here, the $+$ sign is valid at the interface where the reservoir is above the superconductor [the interface at $x=l$ in Figure 4(a) in the main text], and the $-$ sign is valid at the other interface. The Green's function in the reservoir is given by $\underline{\check{g}}$. In the normal state, $\underline{\hat{g}}^R = \hat{\rho}_4$ and the distribution function is given by eq. \eqref{eq:disfnc for a standard voltage}. The average interface magnetization is described by the unit vector $\boldsymbol{m}$, and it enters the magnetization matrix $\check{m}$ as $\check{m} = \hat{1}\otimes \text{diag}(\boldsymbol{m}\cdot \boldsymbol{\sigma}, \boldsymbol{m}\cdot \boldsymbol{\sigma}^*)$. The difference between $\check{m}$ and $\check{m}'$ is that they refer to the average magnetization felt by a quasiparticle transmitted and reflected through the interface, respectively. We assume that $\check{m} = \check{m}'$. The angle difference $\Delta \theta$ is the angle difference between the vectors $\boldsymbol{m}$ at the interfaces. The conductances in the boundary conditions are interpreted as follows. $G_0$ is the tunneling conductance, $G_1$ is a depairing term, $G_{MR}$ is a magnetoresistive term and $G_{\varphi}$ is the spin-mixing term. Under the assumption that all scattering channels have the same polarization $p$, then 
\begin{equation}
    \frac{G_1}{G_0} = \frac{1 - \sqrt{1-p^2}}{1+\sqrt{1-p^2}}, \quad \frac{G_{MR}}{G_0} = \frac{p}{1+\sqrt{1-p^2}}.
\end{equation}
The spin-mixing term can be regarded as a fitting parameter. 

When the Usadel equation is solved, we can calculate observables such as the charge current,
\begin{equation} \label{eq: charge current}
    I = I_0 \int_{0}^{\infty} \text{Re Tr}\left[\hat{\rho}_4\left(\check{g} \frac{\partial \check{g}}{\partial (x/\xi)} \right)^K \right]\text{d}\left(\frac{\varepsilon}{\Delta_0}\right),
\end{equation}
and the superconducting order parameter,
\begin{equation}\label{eq:gap equation}
    \frac{\Delta}{\Delta_0} = -\frac{N_0\lambda}{4}\int_{-\omega_c}^{\omega_c} \hat{g}^K_{23}\left(\frac{\varepsilon}{\Delta_0}\right)\text{d}\left(\frac{\varepsilon}{\Delta_0}\right).
\end{equation}
The prefactor is $I_0=AeN_0\Delta_0^2\xi/8\hbar<0$, with $A$ being the interfacial contact area. The subscript 23 refers to the matrix element at row 2 and column 3. The voltage induced by a temperature gradient is found by solving the root-finding problem $I(\Delta V)=0$. For eq. \eqref{eq:gap equation} to be consistent, the coupling constant $\lambda$ is related to the cutoff energy by $\omega_c = \text{cosh}(1/N_0\lambda)$. $N_0$ is the normal-state density of states at the Fermi level. The order parameter enters the Usadel equation as a fixed self-energy, but it can also be calculated from the Green's function. Since these values should be consistent, the Usadel equation must be solved by fixed-point iterations. 
To this end, we perform fixed-point iterations for both the magnitude $|\Delta|$ of the order parameter $\Delta=|\Delta|\text{e}^{i\phi}$ and its phase gradient $\partial_x\phi$, which both are physical observables. In principle, the fixed-point iterations could be performed on the phase $\phi$ instead of its gradient. However, in the presence of both a thermal and electrical gradient, $\phi$ drifts, meaning that a constant value is added to the phase at all positions at each fixed-point iteration. While this does not change physical observables, which are gauge-invariant, it renders numerical convergence problematic. The convergence criteria for the order parameter are chosen to be
\begin{equation}
    \frac{||\Delta|_n - |\Delta|_{n-1}|}{|\Delta|_n} < 10^{-5}, \quad \frac{|\partial_x \phi_n - \partial_x \phi_{n-1}|}{ \partial_x \phi_{n}} < 10^{-2},
\end{equation}
where $n$ refers to iteration number.

\begin{figure}
    \centering
    \includegraphics[]{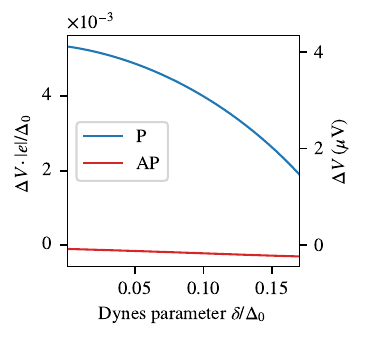}
    \caption{The dependence of the thermovoltage on the Dynes parameter in the P and AP configuration.}
    \label{fig:dynes}
\end{figure}

In the numerical simulations, we use the Dynes parameter $\delta/\Delta_0 = 0.01$. Experimental works have reported significantly higher Dynes parameters in V \cite{shimada2014characterization, ligato2017high}, but with a different growth method. As higher $\delta$ smears out the density of states and reduces the thermoelectric effect as shown in Figure \ref{fig:dynes}, our choice gives a qualitatively better match with the experimental thermovoltage measurements.

\section{Temperature dependence of the thermoelectric effect}
\begin{figure}
    \centering
    \includegraphics{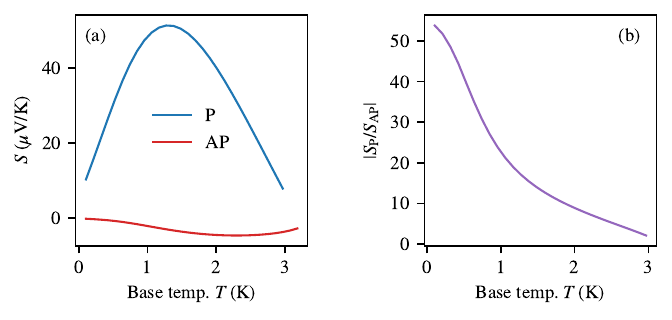}
    \caption{(a) Seebeck coefficient in the P and AP configuration. (b) The ratio between the Seebeck coefficient in the P and AP configuration.}
    \label{fig:tempdep}
\end{figure}
We use the quasiclassical Green's function theory to determine numerically how the thermoelectric effect in the P and AP state depend on the base temperature $T$ of the system. The result is shown in Figure \ref{fig:tempdep}. The parameters used are the same as in the paper, except for the base temperature.

Figure \ref{fig:tempdep}(a) shows the Seebeck coefficient in the P and AP magnetic configurations. We observe that the Seebeck coefficient remains large when the temperature is below the critical temperature of the film.  As the base temperature is increased from a low temperature, the Seebeck coefficient increases. The Seebeck coefficient then reaches its maximum before decreasing for higher temperatures. We note that the Seebeck coefficient is smaller than the experimentally measured value, as discussed in the paper. Figure \ref{fig:tempdep}(b) shows that the the Seebeck coefficient in the P state remains more than an order of magnitude greater than in the AP configuration for temperatures below 2 K $\approx 0.4 T_c^B$. This means that the thermoelectricity do not get spoiled at temperatures well below the critical temperature of the film.


\twocolumngrid


\bibliography{references.bib}